# A modeling of the carbon-nitrogen cycle transport at Igapó I Lake - Londrina, Paraná, Brazil

# Uma modelagem do transporte do ciclo carbono-nitrogênio no lago Igapó I - Londrina, Paraná, Brasil

Suellen Ribeiro Pardo[1]; Paulo Laerte Natti[2]; Neyva Maria Lopes Romeiro[3]; Eliandro Rodrigues Cirilo[4]

[1] Graduada em Matemática, Universidade Estadual Paulista; Mestre em Matemática Aplicada e Computacional, Universidade Estadual de Londrina; suellenrpardo@yahoo.com.br
[2] Docente do Departamento de Matemática da Universidade Estadual de Londrina; plnatti@uel.br
[3] Docente do Departamento de Matemática da Universidade Estadual de Londrina; nromeiro@uel.br
[4] Docente do Departamento de Matemática da Universidade Estadual de Londrina; ercirilo@uel.br

## Abstract

This work is a contribution to better understand the effect that domestic sewage discharges may cause in a water body, specifically Igapó I Lake, in Londrina, Paraná, Brazil. The simulation of the dynamics of pollutant concentrations all over the water body is conducted by means of structured discretization of the geometry of Igapó I Lake, together with the finite differences and the finite elements methods. Firstly, the hydrodynamic flow (without the pollutants), modeled by Navier-Stokes and pressure equations, is numerically resolved by the finite differences method, and associated with the fourth order Runge-Kutta procedure. After that, by using the hydrodynamic field velocity, the flow of the reactive species (pollutants) is described through a transport model, which considers advective and diffusive processes, as well as through a reactions model, restricted to the carbon-nitrogen cycle. The transport and reactions model is numerically resolved by the stabilized finite elements method, by means of a semi-discrete formulation. A qualitative analysis of the numerical simulations conducted in function of the diffusion coefficient provided better understanding of the dynamics of the processes involved in the flow of the reactive species, such as the dynamics of the nitrification process, of the biochemical requirement of oxygen and of the level of oxygen dissolved in the water body at Igapó I Lake.
**Key words:** Water Quality Model. Igapó I Lake. Transport and Reactions of Pollutants. Carbon-Nitrogen Cycle. Finite Elements Method.

## Resumo

Este artigo é uma contribuição para um melhor entendimento do efeito que uma descarga de esgoto doméstico pode causar num corpo d'água, em particular, no lago Igapó I, Londrina, Paraná, Brasil. A simulação da dinâmica das concentrações dos poluentes em todo o corpo d'água é realizada por meio de uma discretização estruturada da geometria do lago Igapó I, juntamente com os métodos de diferenças finitas e de elementos finitos. Primeiramente, o escoamento hidrodinâmico (sem os poluentes), modelado pelas equações de Navier-Stokes e de pressão, é resolvido numericamente pelo método de diferenças finitas, associado ao procedimento de Runge-Kutta de quarta ordem. Em seguida, utilizando o campo de velocidades hidrodinâmico, descreve-se o escoamento das espécies reativas (poluentes) por meio de um modelo de transporte, que considera processos advectivos e difusivos, e por meio de um modelo de reações, restrito ao ciclo carbono-nitrogênio. O modelo de transporte e reações é resolvido numericamente pelo método de elementos finitos estabilizados, através de uma formulação semi-discreta. Uma análise qualitativa das simulações numéricas realizadas, em função do coeficiente de difusão, proporcionou uma melhor compreensão da dinâmica dos processos envolvidos no escoamento de espécies reativas, tais como a dinâmica do processo de nitrificação, da demanda bioquímica de oxigênio e do nível de oxigênio dissolvido no corpo d'água do lago Igapó I.
**Palavras-chave:** Modelo de Qualidade de Água. Lago Igapó I. Transporte e Reações de Poluentes. Ciclo Carbono-Nitrogênio. Método de Elementos Finitos.

# 1 Introduction

The growing demographic and industrial expansion that has been observed in the last decades brought about, as a consequence, hydric pollution caused, among other factors, by the discharge of industrial and domestic sewage. With waters from rivers, lakes and reservoirs compromised, sophistication in the treatment of such a resource is more and more required. Therefore, to understand this issue and search for solutions is a highly important current problem, and one way of solving it is to analyze the relationship between the polluting sources and their degradation mechanisms by using water quality models.

According to Chapra (1997), the history of water quality modeling can be presented in four phases. The first phase had as its landmark the model proposed by Streeter e Phelps (1925). This model described the consumption process of oxygen and the reaeration capacity of the water body by means of two first order ordinary differential equations, considering permanent and uniform flow. Due to lack of computer tools, the models from the 1920's to the 1960's were one-dimensional and limited to the primary treatment of effluents in streams or estuaries with linear kinetics and simple geometries. Such models presented analytical solutions.

In the second phase, during the 1960's, technological advances allowed numerical approaches in more complex geometries. The focus started to be the primary and secondary treatment of effluents and the transport of pollutants in streams and estuaries, modeled in two dimensions. In this period, based on O'Connor e Dobbins's (1958) proposal, models were proposed which consisted of second order differential equations which added the benthonic and photosynthesis demand treatment to the models from the first period.

In the third phase, in the 1970's, the water body started to be observed as a whole. The eutrophyzation process, excessive proliferation of algae caused by nutrients in excess, was the focus of the models. Therefore, representations of the biological processes started to be studied in streams, lakes and estuaries. Simultaneously, non-linear kinetics and three-dimensional models also started to be studied by means of numerical simulations. At that time, concern with the environment and ecological movements increased in some sectors of the society. In this context, in 1971, the Texas Water Development Board (TWDB) created a one-dimensional Water Quality Model (QUAL-I), which allowed the description of advective and diffusive/dispersive transport of pollutants in water bodies (TWDB, 1971). Lately, the United States Environmental Protection Agency (USEPA) improved QUAL-1 model, which started to be called QUAL-II, simulating up to 13 species of parameters indicative of water quality in deeper rivers (ROESNER; GIGUERE; EVENSON, 1981).

The fourth phase ranges from the 1980's to the present. In the beginning of the 1980's there was the emergence of the Task Group on River Water Quality (TGRWQ), a group of scientists and technicians organized by the International Association on Water Quality (IAWQ), which standardized the existing models and manuals. In 1987, due to the several modifications in the QUAL-II model, it was renamed QUAL2E, simulating up to 15 species, accepting punctual and non-punctual sources and fluids both in permanent and non-permanent regime (BROWN; BARNWELL, 1987). The QUAL2K model (CHAPRA; PELLETIER; TAO, 2007) is the current improved version of the QUAL2E model. Parallel to that, in 1985, the USEPA developed the Water Analysis Simulation Program (WASP), which simulated one-dimensional, two-dimensional and three-dimensional processes of conventional and toxic pollutants. This model was also modified several times and its current version is the WASP7 (AMBROSE; WOOL; MARTIN, 2006). Other numerous water quality models can be found in the literature, such as the Activated Sludge Models (ASM1, ASM2 e ASM3), developed by IAWQ (HENZE et al., 2000); the River Water Quality Models (RWQM), also developed by IAWQ (SHANAHAN et al., 2001), the Hydrological Simulation Program-Fortran models (HSPF), developed by USEPA, (BICKNELL et al., 2001), among others.

In this context, this paper is a contribution to better understand the effect that domestic sewage discharge may cause in the water body of Igapó I Lake, located in Londrina, Paraná, Brazil. In order to simulate the effects of such a discharge, a horizontal two-dimensional model (2DH model) is used, in which water flow in the discretized geometry of the lake is described by Navier-Stokes and pressure equations, whereas the transport of the reactive species is described by an advective-diffusive model. Finally, the model WASP6 (AMBROSE; WOOL; MARTIN, 2001) is used in its linear version to describe the reactions of the carbon-nitrogen cycle that occur during the transport of reactive species by the hydrodynamic flow.

This paper is organized as follows. In section 2 the structured discretization that generates the geometry grid of Igapó I Lake is described. The water quality model of our study, consisted of the hydrodynamic model and of the transport and reactions model, is presented in section 3. In section 4 the numerical simulations that provide the local concentrations of the carbon-nitrogen cycle are presented. At last, in section 5, a qualitative analysis of the numerical results, in function of the diffusion coefficient, is conducted.

# 2 Modeling of the geometry at Igapó I Lake

Igapó Lake, located in Londrina, Paraná, Brazil, is situated in the microbasin of Cambé Stream, whose spring

is in the town of Cambé, approximately 10 km from the city of Londrina, in the State of Paraná. After the spring, it flows to the west crossing all the southern area of Londrina, gathering many streams along its way. The Lake is subdivided into: Igapó I, II, III and IV. They were designed in 1957 as a solution for the Cambé Stream drainage problem.

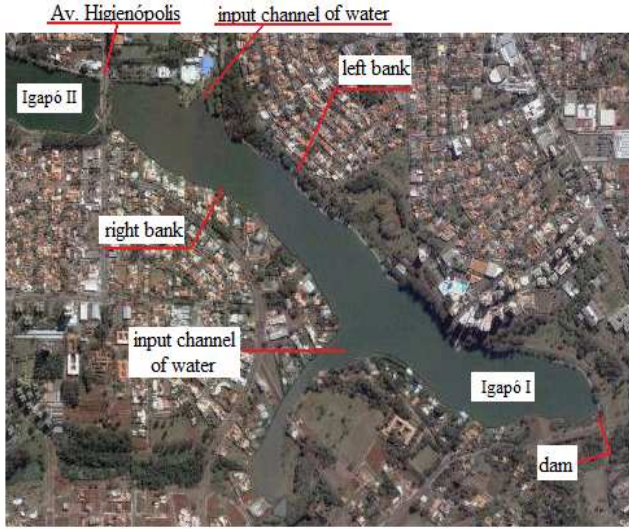

Figure 1: Physical domain of Igapó I Lake.

Because it is located near the central area of the city of Londrina, Igapó I Lake receives the discharge of non-treated pollutants in its waters, besides the discharge of pollutants from Lakes IV, III and II, which pollute Lake I. As observed in Figure 1, the water flows from Igapó II Lake into Igapó I Lake when crossing Higienópolis Avenue, which characterizes its entrance. In the left bank there is undergrowth, as well as an input channel. The right bank is split in private properties and contains another input channel. The exit is a physical dam and the water flow is controlled by water pipes and ramps.

The longitudinal length of the Lake is approximately 1.8 kilometers, while its average width is 200 meters and its average depth is 2 meters. These geophysical characteristics allow modeling the water flow at Igapó I Lake by means of a laminar model type 2DH.

In order to generate the interior grid of Igapó I Lake geometry, a system of elliptical partial differential equations was used, whereas the margins were obtained by parametrized cubic spline polynomial interpolation (CIRILO; BORTOLI, 2006, ROMEIRO; CIRILO; NATTI, 2007). Such procedures were adopted due to their computer performance and due to the similarity obtained between the Lake's geophysical geometry (Figure 1) and the computer grid (Figure 2). It should be observed that in the computer grid in Figure 2, the input channels have been deleted.

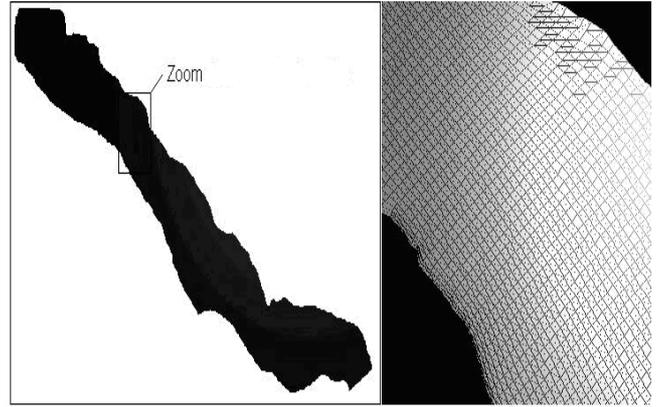

Figure 2: Computer grid of Igapó I Lake.

## 3 Water quality model

The bases for the water quality models are the equations of movement quantity conservation and of continuity (hydrodynamic model), the equations of mass conservation (advective-diffusive transport model) and the equations of reactive processes (reactions model). By means of such equations it is possible to represent the flow dynamics of a water body and study the behavior of the reactive species (pollutants or substances) during transport.

### 3.1 Hydrodynamic model

The water body flow of Igapó I Lake presents laminar characteristics, so that a hydrodynamic model of the type 2DH represents it appropriately. Considering that the fluid (water body) is incompressible, of the Newtonian type and in hydrostatic equilibrium, so the variations of density $\rho$ are not significant (SHAKIB; HUGHES; JOHAN, 1991). Admitting as well that the forces of the external field (actions of wind, heat, ...) are not expressive, that there are no variations in the Lake's contour and using Stokes' hypothesis for Newtonian fluids (SCHLICHTING; GERTEN, 2000, FOX; MCDONALD; PRITCHARD, 2006), the equations of the hydrodynamic model 2DH are given by

$$\frac{\partial u_1}{\partial t} + u_1 \frac{\partial u_1}{\partial x_1} + u_2 \frac{\partial u_1}{\partial x_2} = -\frac{\partial p}{\partial x_1} + \frac{1}{\text{Re}} \left( \frac{\partial^2 u_1}{\partial x_1^2} + \frac{\partial^2 u_1}{\partial x_2^2} \right) \quad (1)$$

$$\frac{\partial u_2}{\partial t} + u_1 \frac{\partial u_2}{\partial x_1} + u_2 \frac{\partial u_2}{\partial x_2} = -\frac{\partial p}{\partial x_2} + \frac{1}{\text{Re}} \left( \frac{\partial^2 u_2}{\partial x_1^2} + \frac{\partial^2 u_2}{\partial x_2^2} \right) \quad (2)$$

$$\nabla^2 p = -\frac{\partial^2 u_1^2}{\partial x_1^2} - 2\frac{\partial^2 u_1 u_2}{\partial x_1 x_2} - \frac{\partial^2 u_2^2}{\partial x_2^2} + \frac{\partial p}{\partial t} + \frac{1}{\text{Re}} \left( \frac{\partial^2 d}{\partial x_1^2} + \frac{\partial^2 d}{\partial x_2^2} \right). \quad (3)$$

Navier-Stokes equations (1) - (2) and pressure equation (3) are in their dimensionless forms (GRESHO; SANI, 1987). They describe the horizontal two-dimensional movement of incompressible Newtonian fluids, establishing the changes in the moments and in the acceleration of the fluid as a result of changes in pressure and in dissipative viscous forces (shear stress), acting in the interior of the fluid.

In equations (1)-(3) the independent variable $t$ is time, $u_1$ and $u_2$ are the components of the velocity vector in longitudinal direction $x_1$ and transversal direction $x_2$, respectively, $p$ is pressure, $d$ is the divergent defined by $d = \frac{\partial u_1}{\partial x_1} + \frac{\partial u_2}{\partial x_2}$ and Re is Reynolds number (FOX; MCDONALD; PRITCHARD, 2006).

## 3.2 Transport and reactions model

In lakes that do not present high concentration of sediments in suspension, the reactive species that are dissolved in the water body flow with the same velocity field of the lake. In these situations, the reactive species are said to be in passive regime, and the study of their transport can be carried out independently of the hydrodynamic modeling.

Considering the mass conservation principle for the reactive species concentration, that the flow variation is almost linear and that the total velocity of the reactive species can broken down into advective velocity $u_i$, for $i = 1,2$ (given by the hydrodynamic model) and into diffusive velocity, which may be modeled by means of Fick's Law, so the transport and reactions model is given by (ROSMAN, 1997)

$$\frac{\partial C}{\partial t} = -u_i \frac{\partial C}{\partial x_i} + D \frac{\partial^2 C}{\partial x_i^2} + \sum R_C,\qquad(4)$$

where it is supposed that the molecular diffusion $D$ of all the reactive species are equal and constant.

In (4), the reactions term $\sum R_C$ can be modeled by means of numerous water quality models, some of them mentioned in the introduction. In this work the model WASP6 (AMBROSE; WOOL; MARTIN, 2001) was used in its linearized version (ROMEIRO; CASTRO; LANDAU, 2003), limited to carbon (C) and nitrogen (N) cycles. In this model the reactions scheme of the carbon (C) and nitrogen (N) cycles is described in Figure 2, by red and blue arrows, respectively.

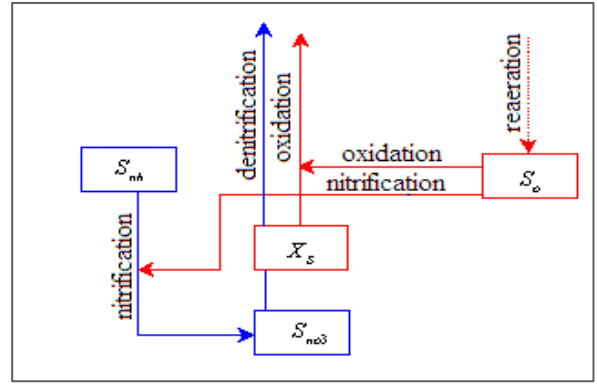

Figure 3: Reactions scheme of carbon (C) and nitrogen (N) cycles, in red and blue arrows, respectively.

In the scheme in Figure 3 the concentrations of the four reactive species under consideration, the concentration of ammonium $S_{nh}$, the concentration of nitrite+nitrate $S_{no3}$, the concentration of the biochemical oxygen demand $X_S$ and the concentration of dissolved oxygen $S_o$, are affected by the C-N cycle processes of the WASP6 model. In the linearized version of the WASP6 model, for carbon and nitrogen cycles (ROMEIRO; CASTRO; LANDAU, 2003), the concentrations of the reactive species are described by the EDO's coupled linear EDO's system below:

$$\begin{cases} \dfrac{dS_{nh}}{dt} = -K_1 S_{nh} - K_7 S_0 + \tau_{S_{nh}} \\[6pt] \dfrac{dS_{no3}}{dt} = K_1 S_{nh} - K_2 S_{no3} + K_8 S_0 + \tau_{S_{no3}} \\[6pt] \dfrac{dX_S}{dt} = -\dfrac{20}{7} K_2 S_{no3} - K_5 X_S - K_4 S_0 + \tau_{X_S} \\[6pt] \dfrac{dS_0}{dt} = -\dfrac{32}{7} K_1 S_{nh} - K_5 X_S - K_6 S_0 + \tau_{S_0}, \end{cases} \qquad (5)$$

where the following constants are defined

$$K_1 = k_{12}\,\Theta_{12}^{T-20}\left(\frac{\overline{S}_0}{k_{nit}+\overline{S}_0}\right)$$

$$K_2 = k_{2D}\,\Theta_{2D}^{T-20}\left(\frac{k_{no3}}{k_{no3}+\overline{S}_0}\right)$$

$$K_3 = k_D \, \Theta_D^{T-20} \left( \frac{\overline{X}_S \, k_{DBO}}{(k_{DBO} + \overline{S}_0)^2} \right)$$
$$+ \frac{64}{14} k_{12} \, \Theta_D^{T-20} \left( \frac{k_{nit} \, \overline{S}_{nh}}{(k_{nit} + \overline{S}_0)^2} \right)$$

$$K_4 = k_D \, \Theta_D^{T-20} \left( \frac{\overline{X}_S \, k_{DBO}}{(k_{DBO} + \overline{S}_0)^2} \right)$$
$$- \frac{32}{7} k_{2D} \, \Theta_D^{T-20} \left( \frac{k_{no3} \, \overline{S}_{no3}}{(k_{no3} + \overline{S}_0)^2} \right)$$

$$K_5 = k_D \, \Theta_D^{T-20} \left( \frac{\overline{S}_0}{k_{DBO} + \overline{S}_0} \right)$$

$$K_6 = k_2 \, \Theta_D^{T-20} + k_D \, \Theta_D^{T-20} \left( \frac{\overline{X}_S \, k_{DBO}}{(k_{DBO} + \overline{S}_0)^2} \right)$$
$$+ \frac{32}{7} k_{12} \, \Theta_D^{T-20} \left( \frac{\overline{S}_{nh} \, k_{nit}}{(k_{nit} + \overline{S}_0)^2} \right)$$

$$K_7 = k_{12} \, \Theta_{12}^{T-20} \left( \frac{\overline{S}_{nh} \, k_{nit}}{(k_{nit} + \overline{S}_0)^2} \right)$$

$$K_8 = k_{12} \, \Theta_{12}^{T-20} \left( \frac{\overline{S}_{nh} \, k_{nit}}{(k_{nit} + \overline{S}_0)^2} \right)$$
$$+ k_{2D} \, \Theta_{12}^{T-20} \left( \frac{k_{no3} \, S_{no3}}{(k_{no3} + \overline{S}_0)^2} \right)$$

$$\tau_{S_{nh}} = K_7 \, \overline{S}_0$$
$$\tau_{S_{no3}} = K_8 \, \overline{S}_0$$
$$\tau_{X_S} = K_4 \, \overline{S}_0$$
$$\tau_{S_0} = k_2 \, \Theta_D^{T-20} \, S_{Sat} + K_3 \, \overline{S}_0 \,,$$

(6)

with $\overline{S}_0$, $\overline{X}_S$, $\overline{S}_{no3}$ and $\overline{S}_{nh}$ the center around which the linearization by Taylor's series was made. The symbols, values and units of the parameters of the linearized reactions model (5-6) are given in Table 1.

Table 1: Symbols, values and units of the constants of the WASP6 model, at steady temperature of 20 °C (AMBROSE; WOOL; MARTIN, 2001).

| Symbol | Value | Unit | Parameter |
|---|---|---|---|
| $\Theta_{2D}$ | 1.045 | | Temperature coefficient for denitrification. |
| $\Theta_{12}$ | 1.08 | | Temperature coefficient for nitrification. |
| $\Theta_D$ | 1.047 | | Temperature coefficient for Carbon oxydation. |
| $\Theta_2$ | 1.028 | | Temperature coefficient for reaeration. |
| $k_{2D}$ | 0.09 | $h^{-1}$ | Denitrification index. |
| $k_{12}$ | 0.22 | $h^{-1}$ | Nitrification index. |
| $k_D$ | 0.38 | $h^{-1}$ | Oxydation index. |
| $k_2$ | 1.252 | $h^{-1}$ | Reaeration index. |
| $k_{DBO}$ | 0.001 | $mgL^{-1}$ | Half saturation constant of the carbonated Biochemical Oxigen Demand. |
| $k_{nit}$ | 0.2 | $mgL^{-1}$ | Half saturation constant for Dissolved Oxigen limited to the nitrification process. |
| $k_{no3}$ | 0.1 | $mgL^{-1}$ | Half saturation constant for Dissolved Oxigen limited to the denitrification process. |
| $S_{Sat}$ | 8.3 | $mgL^{-1}$ | Dissolved Oxigen saturation concentration. |

Substituting the reactions model (5-6) in (4), for the four reactive species under consideration, we will have the transport and reactions model, that is,

$$\frac{\partial S_{nh}}{\partial t} + u_i \frac{\partial S_{nh}}{\partial x_i} - D \frac{\partial^2 S_{nh}}{\partial x_j^2} = -K_1 S_{nh} - K_7 S_0 + \tau_{S_{nh}}$$

$$\frac{\partial S_{no3}}{\partial t} + u_i \frac{\partial S_{no3}}{\partial x_i} - D \frac{\partial^2 S_{no3}}{\partial x_j^2} = K_1 S_{nh} - K_2 S_{no3} + K_8 S_0 - \tau_{S_{no3}}$$

(7)

$$\frac{\partial X_S}{\partial t} + u_i \frac{\partial X_S}{\partial x_i} - D \frac{\partial^2 X_S}{\partial x_j^2} = -\frac{20}{7} K_2 S_{no3} - K_5 X_S - K_4 S_0 + \tau_{X_S}$$

$$\frac{\partial S_0}{\partial t} + u_i \frac{\partial S_0}{\partial x_i} - D \frac{\partial^2 S_0}{\partial x_j^2} = -\frac{32}{7} K_1 S_{nh} - K_5 X_S - K_6 S_0 + \tau_{S_0}$$

where the indexes $i=1,2$ represent the longitudinal and transversal directions, respectively, in relation to the computer grid lines, $u_i$ are the components of the vector velocity given by the hydrodynamic model and $D$ is the diffusion coefficient of the reactive species.

## 4  Numerical simulations

For the numerical simulations, it is supposed that Igapó I Lake does not present sources and drains, except for the entrance and exit dams, as modeled in Figure 2.

To generate the interior of Igapó I Lake geometry, a structured discretization was used, in generalized coordinates, by means of an elliptical PDE's system, whereas the margins were obtained by means of parameterized cubic spline polynomial interpolation (CIRILO; DE BORTOLI, 2006). Such procedures were adopted due to their computer performance as well as by the fast similarity obtained with the physical geometry, from little known points of the domain (MALISKA, 1995). The study considered 839 points located along the left and right margins and 35 points located in the entrance and exit contours. To solve the resulting tridiagonal linear systems (RUGGIERO; LOPES, 1996), the TDMA (TriDiagonal Matrix Algorithm) procedure was utilized, which considerably reduces memory time (DE BORTOLI, 2000).

In the numerical resolution of the PDE's systems (1-3), that constitute the horizontal two-dimensional hydrodynamic model, the block technique was used (DE BORTOLI, 2000). In this procedure, firstly, it is necessary to conduct an analysis of which and how many subgrids (sub-blocks) would constitute the whole grid. After distinguishing the sub-blocks, for each one, it is necessary to define the contour and their interior. Finally, the sub-blocks are read and recorded in files referring to the whole grid. The connection of the sub-blocks that compose the grid is made from the reading of the extreme points common among the sub-blocks (CIRILO; DE BORTOLI, 2006). So, the PDE's systems (1-3) were resolved in generalized coordinates, approaching the terms of the spatial derivates according to central differences. For the discretized velocity field, the explicit fourth order Runge-Kutta method was used and for the discretized pressure field, the Gauss-Seidel method, with successive relaxations, was used (SMITH, 1985).

As for the numerical resolution of the transport model (7), the stabilized finite elements method was employed, in its Galerkin's semi-discrete formulation, where the spatial derivates are approached by finite elements and temporal derivates are approached by finite differences (ROMEIRO; CASTRO, 2007). A stabilization procedure of the Streamline Upwind Petrov-Galerkin (SUPG) type, proposed by BROOKS and HUGHES (1982), was also employed. Recently, several attempts to extend the SUPG method have been developed to account for incompressible and compressible flows (COSTA; LYRA; LIRA, 2005).

About the value of the diffusion coefficient $D$, in (7), it is supposed that it is constant throughout Igapó I Lake geometry and equal for all the reactive species. According to (CHAPRA, 1997), the several reactive species have molecular diffusion values in the interval between $D=10^{-3} m^2 h^{-1}$ and $D=10^{-1} m^2 h^{-1}$. However, the turbulent diffusion coefficient in rivers and lakes, which depends on the turbulent phenomenon scale, takes values between $D=10^1 m^2 h^{-1}$ and $D=10^{10} m^2 h^{-1}$. With the objective of simulating and studying in our model the turbulent and molecular diffusion phenomena, involved in the transport of the reactive species at Igapó I Lake, the following values for the diffusion coefficient will be taken: $D=10^{-3} m^2 h^{-1}$, $D=1\ m^2 h^{-1}$ and $D=10^4 m^2 h^{-1}$. Therefore, by means of numerical simulations of system (7), in function of the diffusion coefficient $D$, the qualitative aspects of the carbon and nitrogen cycles transport in the water body of Igapó I Lake, Londrina, Paraná, will be studied. It is highlighted that this numerical simulation is not aimed at providing quantitative predictions about the pollution index in a particular place of the Lake's physical domain at a certain time. It is known that the entrance conditions (initial and boundary) of the reactive species vary daily. In fact, measurements that have been conducted since 2007 (TI SOLUTION, 2010) by a joint cooperation involving the Environmental Institute of Paraná (IAP), the Municipal Council for the Environment of Londrina (CONSEMMA), the State University of Londrina (UEL) and the Engineering and Architecture Club of Londrina (CEAL), provided very discrepant measurements of the Water Quality Index (WQI) and individual measurements of the water quality parameters, according to collecting date. Under such conditions, our objective is to provide qualitative information, such as the most polluted places in the Lake's domain, independently of initial concentrations and of reactive species boundary.

In this context, by using the mathematical model developed, this work will describe qualitatively the impact that continuous ammonium discharge at Igapó I Lake entrance has throughout its extension, characterized by the area between Higienópolis Avenue and the dam (Figure 1).

Firstly, there is the numerical calculation, from the hydrodynamic model (1-3), of the components of the vectorial field of velocities $\vec{u}=(u_1,u_2)$ and of the scalar field of pressures $p$, in all of Igapó I Lake domain. In this procedure, Reynolds number $Re=10$ was used and initial and boundary conditions below.

- Initial conditions of the hydrodynamic model.

It is considered that at the initial moment $t = 0$ the velocity and pressure fields, in the interior points of Igapó I Lake geometry, are given by

$$\vec{u}(X,0) = (1,0) \text{ and } p(X,0) = 1.0 \quad , \tag{8}$$

where $X = (x_1, x_2)$ is an interior point of the grid domain at Igapó I Lake, see Figure 2. For $X = \overline{X}$, entrance points of Igapó I Lake, and for $X = \tilde{X}$, exit points of Igapó I Lake, see Figures 1 and 2, the following values for initial velocities and pressures are taken,

$$\begin{aligned}\vec{u}(\overline{X},0) = (1,0) &\quad \text{and} \quad p(\overline{X},0) = 1.0 \\ \vec{u}(\tilde{X},0) = (1,0) &\quad \text{and} \quad p(\tilde{X},0) = 1.0 \end{aligned} \tag{9}$$

Finally, for the other points of Igapó I Lake boundary, the following conditions are considered

$$\vec{u}(X_p,0) = (0,0) \quad \text{and} \quad p(X_p,0) = 0.0 \quad , \tag{10}$$

where $X_p$ are points of the Lake's margin, except the entrance and exit ones.

- Boundary conditions of the hydrodynamic model.

It is considered that the velocity field, for $t > 0$, in the entrance and exit points of Igapó I Lake geometry is free, that is, it is determined by the numerical problem. In the other points of the margin it is null,

$$\begin{aligned}\vec{u}(\overline{X},t) &\quad \text{free condition} \\ \vec{u}(\tilde{X},t) &\quad \text{free condition} \\ \vec{u}(X_p,t) &= (0,0) \quad . \end{aligned} \tag{11}$$

In relation to the pressure field, for $t > 0$, it is considered a gradient of 10 % between entrance pressure and exit pressure of Igapó I Lake. In the other points of the margin, the pressures are calculated by numerical procedure, so

$$\begin{aligned}p(\overline{X},t) &= 1.0 \\ p(\tilde{X},t) &= 0.9 \\ p(X_p,t) &\quad \text{free condition} . \end{aligned} \tag{12}$$

From such conditions, the advective velocity field and the pressure field is numerically calculated using the hydrodynamic model (1-3).

In order to simulate the dynamics of the reactive species concentrations in (7), in the modeled geometry of Igapó I Lake, initially the values of the parameters defined in (6) are calculated by using the data in Table 1, that is,

$$\begin{aligned}K_1 &= 9.67 \times 10^{-3} \; h^{-1} \\ K_2 &= 4.67 \times 10^{-5} \; h^{-1} \\ K_3 &= 4.67 \times 10^{-5} \; h^{-1} \\ K_4 &= 1.22 \times 10^{-6} \; h^{-1} \\ K_5 &= 1.66 \times 10^{-2} \; h^{-1} \\ K_6 &= 5.38 \times 10^{-2} \; h^{-1} \\ K_7 &= 4.77 \times 10^{-5} \; h^{-1} \\ K_8 &= 4.77 \times 10^{-5} \; h^{-1} \\ \tau_{S_{nh}} &= 3.96 \times 10^{-4} \; mgL^{-1}h^{-1} \\ \tau_{S_{n03}} &= 3.96 \times 10^{-4} \; mgL^{-1}h^{-1} \\ \tau_{X_S} &= 1.01 \times 10^{-5} \; mgL^{-1}h^{-1} \\ \tau_{S_0} &= 4.47 \times 10^{-1} \; mgL^{-1}h^{-1} \quad . \end{aligned} \tag{13}$$

As for the initial and contour conditions for the transport and reactions model (7), the values below are taken.

- Initial conditions for the transport and reactions model.

It is considered that at the initial time $t = 0$ the scalar field of concentrations of a reactive species is null, that is,

$$C(X_m,0) = 0.00 \; mgL^{-1} \quad , \tag{14}$$

where $X_m = (x_1, x_2)$ are all the grid points (interior and boundary) of Igapó I Lake. Therefore, the initial conditions for the concentrations of ammonium $S_{nh}$, of nitrite+nitrate $S_{n03}$, of the biochemical demand of oxygen $X_S$ and of dissolved oxygen $S_0$ are given by

$$\begin{aligned}S_{nh}(X_m,0) &= 0.00 \; mgL^{-1} \\ S_{no3}(X_m,0) &= 0.00 \; mgL^{-1} \\ X_S(X_m,0) &= 0.00 \; mgL^{-1} \\ S_0(X_m,0) &= 0.00 \; mgL^{-1} \quad . \end{aligned} \tag{15}$$

- Borderline conditions for the transport and reactions model.

For $t > 0$ and $X = \overline{X}$, Igapó I Lake entrance points, the following constant borderline concentrations are considered

$$S_{nh}(\overline{X},t) = 1.74 \; mgL^{-1}$$
$$S_{n03}(\overline{X},t) = 0.00 \; mgL^{-1}$$
$$X_S(\overline{X},t) = 5.05 \; mgL^{-1}$$
$$S_0(\overline{X},t) = 8.30 \; mgL^{-1}$$
(16)

and for $X = \widetilde{X}$, Igapó I Lake exit points, free conditions, that is,

$$S_{nh}(\widetilde{X},t) \quad \text{free condition}$$
$$S_{n03}(\widetilde{X},t) \quad \text{free condition}$$
$$X_S(\widetilde{X},t) \quad \text{free condition}$$
$$S_0(\widetilde{X},t) \quad \text{free condition}.$$
(17)

In the other borderline points, for $t > 0$, null concentrations are taken, due to the hypothesis of absence of other sources and drains, so

$$S_{nh}(X_p,0) = 0.00 \; mgL^{-1}$$
$$S_{n03}(X_p,0) = 0.00 \; mgL^{-1}$$
$$X_S(X_p,0) = 0.00 \; mgL^{-1}$$
$$S_0(X_p,0) = 0.00 \; mgL^{-1}.$$
(18)

It should be noticed that the borderline conditions at the entrance of the lake represent a continuous discharge of 1.74 milligrams of ammonium per liter, absence of nitrite+nitrate concentration, which will be generated by means of the nitrification process, a biochemical demand of 5.05 milligrams of oxygen per liter and a saturation concentration of dissolved oxygen of 8.30 milligrams per liter.

From the advective velocity field, given by the hydrodynamic model (1-3) and by conditions (8-12), it is simulated, from the transport and reactions model (7) and from conditions (13-18), the transport of the concentrations of ammonium, of nitrite+nitrate, of the biochemical demand of oxygen and of the dissolved oxygen, all over the Lake's domain, at all times, for the following diffusion coefficients: $D = 10^{-3} \; m^2 h^{-1}$, $D = 1.0 \; m^2 h^{-1}$ and $D = 10^4 \; m^2 h^{-1}$. Figures (4-6) present the results of the numerical simulations of the four reactive species, in a time interval of 300 hours of continuous discharge, when the flow reaches an almost-stationary situation.

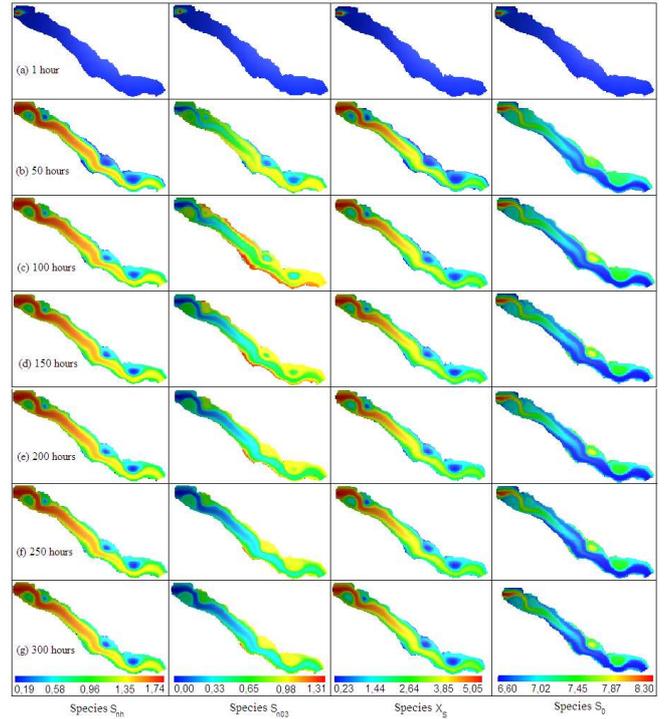

Figure 4: Concentration of the reactive species in Igapó I Lake, in function of time, when $D = 10^{-3} m^2 h^{-1}$.

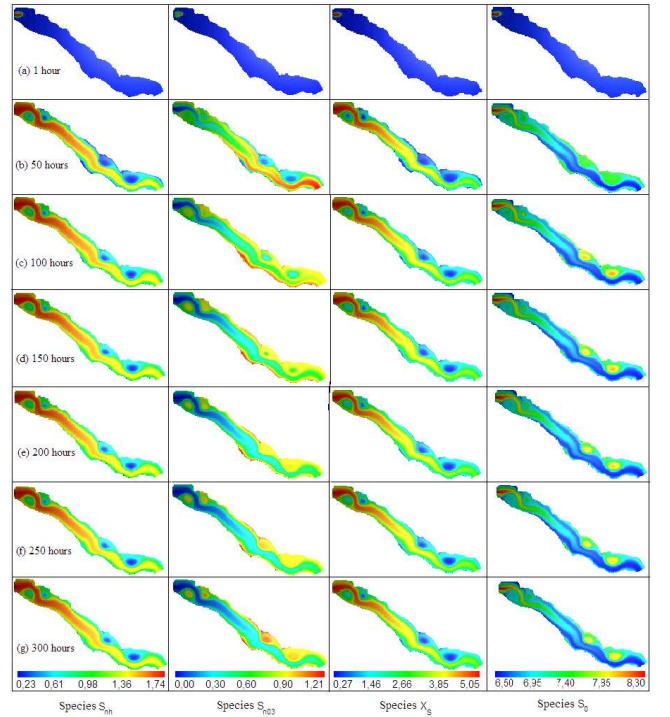

Figure 5: Concentration of the reactive species in Igapó I Lake, in function of time, when $D = 1.0 \; m^2 h^{-1}$.

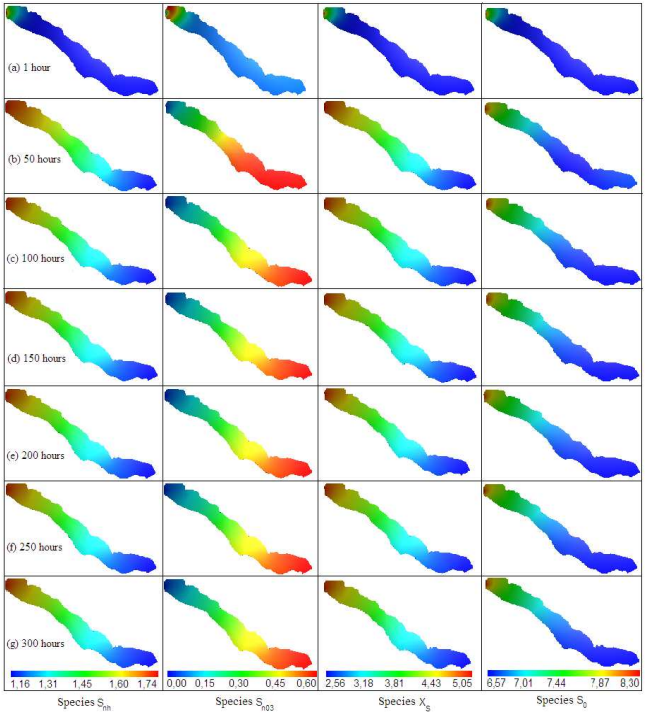

Figure 6: Concentration of the reactive species Igapó I Lake, in function of time, when $D = 10^4 m^2 h^{-1}$.

Below are listed some results that can be observed in the numerical simulations presented in Figures (4-6).

1. The numerical procedure, when $D$ takes values compatible with the molecular diffusion, $D = 10^{-3} m^2 h^{-1}$, captures four big vortices, two of them located near Igapó I Lake entrance, one on the left margin and the other on the right margin of the Lake, and the other two on the left margin, after the central region of the Lake. However, for the turbulent diffusion values, $D = 10^4 m^2 h^{-1}$, a destructuration of this topology of vortices is observed, characterizing the fact that the diffusive velocity field is bigger than the advective velocity field.

2. It is observed that the concentration of ammonium continuously discharged in the Lake's entrance decreases along the flow, generating higher concentration values of nitrite+nitrate in the Lake's exit. Consistently, a high index of biochemical demand of oxygen is observed in the regions with high concentrations of ammonium.

3. It is highlighted that in non-turbulent flow situations, $D = 10^{-3} m^2 h^{-1}$ and $D = 1.0 m^2 h^{-1}$, the higher concentrations of nitrite+nitrate occur in the Lake's vortices, characterizing them as the regions that are most polluted by nitrite and nitrate.

4. Lastly, Figures (4-6) and Table 2 show that for the diffusion coefficients ranging from $D = 10^{-3} m^2 h^{-1}$ a $D = 10^4 m^2 h^{-1}$, the variation intervals of the reactive species concentrations decrease, in agreement with the fact that the effective velocity field (advection+diffusion) has increased. Therefore, it is verified that the higher the flow diffusion coefficient, the lower the Lake's self-depuration capacity.

Table 2: Minimum and maximum values of the reactive species concentrations, in $mgL^{-1}$, considering all the points in the geometry grid of Igapó I Lake, in function of diffusion $D$, in $m^2 h^{-1}$, for $t = 300$ hours.

|  | $S_{nh}$ | $S_{n03}$ | $X_S$ | $S_0$ |
|---|---|---|---|---|
| $D = 10^{-3}$ | 0,19–1,74 | 0,00–1,31 | 0,23–5,05 | 6,60–8,30 |
| $D = 1$ | 0,23–1,74 | 0,00–1,21 | 0,27–5,05 | 6,50–8,30 |
| $D = 10^4$ | 1,16–1,74 | 0,00–0,60 | 2,60–5,05 | 6,57–8,30 |

## 5 Conclusions

In this work, a transport and reactions model was designed in order to simulate the evolution of pollutants discharge in the water body of Igapó I Lake. In the modeling of the problem, a laminar flow for Igapó I Lake was supposed, so that the hydrodynamic model was described by a 2DH model given by Navier-Stokes and pressure equations (1-3). In this hydrodynamic model, it was considered that water is an incompressible Newtonian fluid, and the transport of the reactive species (carbon and nitrogen cycles) occurs passively. In order to describe the reactive processes during the flow, AMBROSE, *et al.* (2001) linearized model was used, which is shown schematically in Figure 3. Finally, the horizontal two-dimensional model of advective-diffusive-reactive transport of the carbon and nitrogen cycles in Igapó I Lake is given by differential equations (7) and by initial and contour conditions (8-18).

The numerical simulations of the problem of initial and contour conditions (7-18), as a function of the diffusion coefficient D, provided better understanding of the dynamics of the processes involved in the reactive species flow, such as the dynamics of the nitrification process, of the biochemical demand of oxygen and of the level of dissolved oxygen in the water body. In the simulations, a continuous discharge of 1.74 milligrams of ammonium per liter in the entrance of Igapó I Lake is considered. Figures (4-6) present the results of the transport simulations of the concentrations of ammonium, of nitrite+nitrate, of the biochemical oxygen demand, and of the dissolved oxygen, all over the Lake's domain, at several times, for the diffusion coefficient values $D = 10^{-3}\ m^2 h^{-1}$, $D = 1\ m^2 h^{-1}$ e $D = 10^4 m^2 h^{-1}$. By analyzing the numerical results presented, it is possible to verify that in the non-turbulent regime, the most relevant fact is the occurrence of high concentrations of nitrite and nitrate in the Lake's vortices, characterizing them as the Lake's most polluted regions. However, for diffusion values in the turbulent regime, $D = 10^4 m^2 h^{-1}$, the results presented in Figure (6) show that the flow of reactive species in the Lake is similar to the flow in a river, with the disappearance of the hydrodynamic vortices. Finally, when $D = 1\ m^2 h^{-1}$, there is flow dynamics compatible with observations conducted at Igapó I Lake by a team from the Laboratory of Simulation and Numerical Analysis (LabSAN) from the State University of Londrina.

## Acknowledgements

The author P. L. Natti thanks the State University of Londrina for the financial support obtained by means of the Programs FAEPE/2005 and FAEPE/2009. The author N. M. L. Romeiro acknowledges National Council for Scientific and Technological Development (CNPq-Brazil) for the financial support to this research (CNPq 200118/2009-9).